\documentclass[final,12pt,twocolumn]{elsarticle}

\usepackage[margin=2.5cm]{geometry}

\usepackage{amssymb}
\usepackage{tikz}
\usetikzlibrary{arrows,shapes}
\usepackage{booktabs} 
\usepackage{float,subfloat} 
\usepackage{graphicx} 
\usepackage{amsmath} 
\usepackage{hyperref}
\usepackage{graphicx}
\usepackage{epstopdf}
\usepackage{subfig}
\usepackage{fancyhdr}
\graphicspath{{figures/}}
\usepackage{bm}
\usepackage[linewidth=2pt]{mdframed}
\usepackage{lipsum}
\usepackage[toc,page]{appendix} 
\usepackage{algorithm}
\usepackage[noend]{algpseudocode}
\usepackage{listings}
\usepackage{color} 
\definecolor{mygreen}{RGB}{28,100,0} 
\definecolor{mylilas}{RGB}{170,55,241}
\usepackage{multicol}
\usepackage[numbered]{matlab-prettifier}
\usepackage{tcolorbox}
\tcbuselibrary{listings,skins,theorems}

\begin{document}

\lstdefinelanguage{Maple}%
{xleftmargin=0.3cm,
basicstyle=\fontsize{8}{9}\selectfont,
morekeywords={and,assuming,break,by,catch,description,do,done,%
elif,else,end,error,export,fi,finally,for,from,global,if,%
implies,in,intersect,local,minus,mod,module,next,not,od,%
option,options,or,proc,quit,read,return,save,stop,subset,then,%
to,try,union,use,uses,while,xor},%
sensitive=true,%
morecomment=[l]\#,%
morestring=[b]",%
morestring=[d]"%
literate=%
{Ã¥}{{\aa}}1
}[keywords,comments,strings]%

\lstset{language=Matlab,%
    xleftmargin=0.3cm,
    basicstyle=\fontsize{8}{9}\selectfont,
    breaklines=true,%
    morekeywords={matlab2tikz},
    morekeywords=[2]{1}, keywordstyle=[2]{\color{black}},
    identifierstyle=\color{black},%
    stringstyle=\color{mylilas},
    commentstyle=\color{mygreen},%
    showstringspaces=false,
    numbers=left,%
    numberstyle={\tiny \color{black}},
    numbersep=3pt, 
    emph=[1]{if,else,for,end,break},emphstyle=[1]\color{blue}, 
}

\makeatletter
\def\ps@pprintTitle{%
 \let\@oddhead\@empty
 \let\@evenhead\@empty
 \def\@oddfoot{}%
 \let\@evenfoot\@oddfoot}
\def\BState{\State\hskip-\ALG@thistlm}
\makeatother

\begin{frontmatter}



\title{SIR - an Efficient Solver for Systems of Equations}


\author{Jan Scheffel}
\author{Kristoffer Lindvall}

\address{Department of Fusion Plasma Physics, School of Electrical 		Engineering \\
KTH Royal Institute of Technology, Stockholm, Sweden}

\begin{abstract}
The Semi-Implicit Root solver (SIR) is an iterative method for globally convergent solution of systems of nonlinear equations. Since publication \cite{Scheffel:SIR}, SIR has proven robustness for a great variety of problems. We here present MATLAB  and MAPLE codes for SIR, that can be easily implemented in any application where linear or nonlinear systems of equations need be solved efficiently. The codes employ recently developed efficient sparse matrix algorithms and improved numerical differentiation. SIR convergence is quasi-monotonous and approaches second order in the proximity of the real roots. Global convergence is usually superior to that of Newton's method, being a special case of the method. Furthermore the algorithm cannot land on local minima, as may be the case for Newton's method with linesearch. \end{abstract}

\begin{keyword}
Newton method \sep Jacobian \sep root solver \sep equation solver \sep MATLAB


\end{keyword}

\end{frontmatter}

\section{Motivation and significance}
\noindent Systems of algebraic equations generally need be solved computationally, whether it be with direct methods or iterative methods. The Semi-Implicit Root solver (SIR \citep{Scheffel:SIR}), reported on here, was developed in order to improve on the global convergence characteristics of the widely used Newton method. Linesearch \cite{NumericalRecipes} is often combined with Newton's method to improve convergence but, unlike SIR, it may lead to local extrema rather than to the roots of the equations. SIR development was initially inspired by semi-implicit PDE algorithms (\cite{Gottlieb:1},\cite{Harned:1},\cite{Harned:2}) and evolved as a robust equation solver for the time-spectral method GWRM \citep{Scheffel:GWRM1} for systems of PDEs; it is however generally applicable to systems of equations.    

Recently the algorithm, and its coding in MATLAB and MAPLE, has been substantially improved with respect to efficiency. Updated software and matrix handling is now used. We believe it would be beneficial to present ready-to-use codes, being the motivation for this paper. First, a brief overview of SIR is provided in section 2. For a thorough explanation of the SIR algorithm the reader is advised to consult \citep{Scheffel:SIR}. An example application is given in section 3. Pseudocode, describing the algorithm in detail, can be found in section 4. In Appendix, MATLAB and MAPLE codes are provided. 

\section{Software description}
\noindent The roots of the \textit{single equation} $f(x) = 0$, with $f(x) \ { \equiv} \ x-\varphi(x)$, are found by SIR after a reformulation in iterative form as
\begin{align}
x^{i+1}+{\beta}x^{i+1}={\beta}x^i+\varphi(x^i),
\label{eq:1}
\end{align}
where $\beta$ is a real and arbitrary parameter. Eq. \ref{eq:1} will have the same roots as the original equation. We cast it into the form
\begin{align}
x^{i+1}=\alpha(x^i-\varphi(x^i))+\varphi(x^i),
\label{eq:2}
\end{align}
where $\alpha=\beta/(1+\beta)$ is a parameter for optimizing global and local convergence. Introducing $\varPhi(x;\alpha) \ { \equiv} \ \alpha(x-\varphi(x)) +\varphi(x)$ it may be shown that convergence requires
$|\partial \varPhi /\partial{x}|<1$ to hold for the iterates $x^i$ in a neighbourhood of the root \citep{Scheffel:SIR}.
Newton's method (for a single equation also known as Newton-Raphson's method) assumes $|\partial \varPhi /\partial{x}|=0$ everywhere, thus potentially achieving maximized, second order convergence near the root.  NewtonÕs method is, however, not globally convergent because of the breakdown of the linear approximation for initial iterates $x^0$ positioned too distant from the root. This problem is remedied by SIR through enforcing monotonic convergence. The trick is to choose appropriate values of $|\partial \varPhi /\partial{x}| \equiv R$ at each iteration $i$. 

Thus SIR iterates the equation
\begin{align}
x^{i+1}=\alpha^i(x^i-\varphi(x^i))+\varphi(x^i),
\label{eq:5}
\end{align}
using
\begin{align}
\alpha^i=\frac{R^i-\varphi^{'}(x^i)}{1-\varphi^{'}(x^i)}.
\label{eq:4}
\end{align}
where $\varphi^{'}(x)={d}\varphi/dx$. Typically $R^{i}$ is initially given a value in the interval $[0.5,0.99]$ whereafter SIR automatically reduces it towards zero to achieve second order convergence in the vicinity of the root.
Since monotonic convergence is guaranteed, SIR will consecutively find all real roots $x$ to the equation.

Generalizing to \textit{systems of equations}, SIR now solves
\begin{align}
\textbf{x}^{i+1}=\textbf{A}^i(\textbf{x}^i-\varphi(\textbf{x}^i))+\varphi(\textbf{x}^i),
\label{eq:6}
\end{align}
where
\begin{align*}
\textbf{A}=\textbf{I}+(\textbf{R}-\textbf{I})\textbf{J}^{-1},
\end{align*}
\begin{align*}
(\textbf{R})_{mn}=\delta_{mn}R_m.
\end{align*}
Here the Jacobian matrix $\textbf{J}$ has components $J_{mn}=\partial(x_m-\varphi_m(\textbf{x}))/\partial{x_n}$, $\delta_{mn}$ is the Kronecker  delta and $\textbf{I}$ is the identity matrix. It is usually most economic to obtain $\textbf{A}$ by solving the linear matrix relation
\begin{align*}
\varPhi^{'}=(\textbf{A}-\textbf{I})\textbf{J}+\textbf{I},
\end{align*}
using $\partial\varPhi_{m}/\partial{x_n}=R_{mn}$ (diagonal matrix) and sparse matrix methods.

When solving multidimensional equations, strict monotonous convergence cannot be guaranteed, since there is no known effective procedure (like root bracketing in the 1D case) to safely maintain the direction from a starting point towards a root. A procedure for "quasi-monotonicity" is thus employed in SIR \citep{Scheffel:SIR}. SIR is also safeguarded against certain pitfalls. An evident problem can be seen in Eq. (4); there is a risk that the $\textbf{A}$ matrix may become singular. The cure for this is subiteration, where $R_{m}$ values are modified towards unity. See \citep{Scheffel:SIR} for further discussion of measures that enhance convergence. 

Summarizing, at each iteration the SIR algorithm reduces $R_{m}$ values towards zero to approach second order convergence. If non-monotonous convergence becomes pronounced in any dimension, local subiteration by increasing $R_{m}$ values towards unity is used.       

\section{Illustrative example}
\noindent SIR has been compared to Newton's method with line-search (NL) for a large set of standard problems \citep{Scheffel:SIR}. In several cases it features superior convergence characteristics, in particular when singularities appear in the computation of the \textbf{A} matrix. 

It is well known that the NL method may sometime land on local minima rather than on the roots of the equations. An example of this was provided in \citep{Scheffel:SIR}, where also the concept of "convergence diagrams" were introduced. Since iterative solvers can be very sensitive to the starting points $x^0$, the diagram displays, using colour marking, the quality of convergence using a set of 61 by 61 uniformly distributed $x^0$ $\in$ [-5,5]. The two equations solved are $x_1=\text{cos}(x_2)$ and $x_2=3\text{cos}(x_1)$.  

The convergence diagrams clearly show that the global convergence properties of the SIR and NL schemes are complex even for this seemingly simple problem. It is seen that, in subiteration mode (SIR-s), convergence is rapid for nearly all starting points, whereas here standard SIR converges for about one third of the starting points. The NL method cannot match the convergence of SIR-s, due to excessive landing on non-zero local minima, and performs similarly as SIR.

Recently we have introduced a number of measures to improve SIR efficiency. An important example concerns the time-consuming computation of the $\textbf{A}$ matrix, which acts as a "scaffold" when building the solution. The scaffold needs not be perfect, however. In nonlinear computations it may suffice to recompute $\textbf{A}$ the first few iterations only (parameter $mA$; see MAPLE code), whereafter it accurately leads to convergence. The fact that the $\textbf{A}$ is a constant for linear systems of equations is also useful. 

The MATLAB implementation has been significantly improved by minimizing the number of symbolic operations performed.  
 
\section{MATLAB code for the example}
\noindent Below follows a MATLAB coding of the example described above.

\begin{lstlisting}
%%%%%%%%%%%%%%%%%%%%%%%%%%%%%%%%%%%%
%% Find roots with SIR and fsolve %%
%%%%%%%%%%%%%%%%%%%%%%%%%%%%%%%%%%%%
clear all
%% 1D function - f(x)=x-2cos(x).
% Find the root x*, 
% SIR phi(x)=x-f(x)=2cos(x).
disp('1D case')
fun1D = @(x)(x-2*cos(x));
x0 = 2;

tic;y1D = fsolve(fun1D,x0);toc
disp('fsolve solution')
disp(y1D)

phi1D = @(x)(2*cos(x));
tic;U1D = SIR(phi1D,x0,0);toc
disp('SIR solution')
disp(U1D)

%% 2D function - f1(x1,x2)=x1-cos(x2),
%                f2(x1,x2)=x2-3cos(x1).
% Find roots (x1*,x2*), 
% SIR  phi1(x1,x2)=x1-f1(x1,x2)=cos(x2),
%      phi2(x1,x2)=x2-f2(x1,x2)=3cos(x1).

disp('2D case')
fun2D = @(x)[x(1)-cos(x(2)),...
    x(2)-3*cos(x(1))];
x0 = [-2,-2];

tic;y2D = fsolve(fun2D,x0);toc
disp('fsolve solution')
disp(y2D)

phi2D = @(x)[cos(x(2)),3*cos(x(1))];

tic;U2D = SIR(phi2D,x0,1);toc
disp('SIR solution')
disp(U2D)

%%     Surfplot    %%
% Z = (f1^2+f2^2)/2 %
hold on
[X,Y] = meshgrid(-5:0.65:5);
f1 = X-cos(Y); f2 = Y-3*cos(X);
Z = (f1^2+f2^2)/2;
s = surf(X,Y,Z,'FaceAlpha',0.45);
view([-37 60])
% Point-plot of initial guess x0
plot3(x0(1),x0(2),0,'Marker','o',...
    'MarkerEdgeColor','r',...
    'MarkerFaceColor','r')
% Point-plot of solution (x1*,x2*)
plot3(U2D(1),U2D(2),((U2D(1)-cos(U2D(2)))^2+...
    (U2D(2)-3*cos(U2D(1)))^2)/2,...
    'Marker','o',...
    'MarkerEdgeColor','b',...
    'MarkerFaceColor','b')
hold off
print -depsc surfFig
\end{lstlisting}
\begin{figure}[H]
    \centering
    \includegraphics[height=2.8in,width=3in]{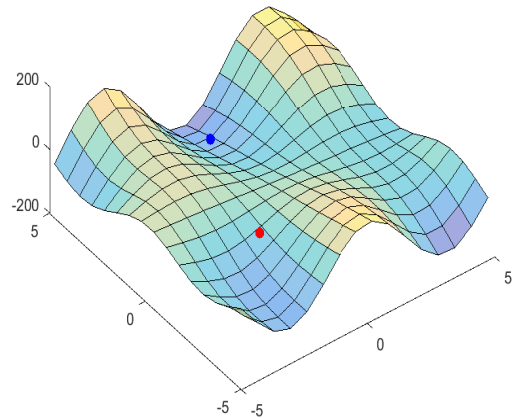}
    \caption{Surfplot of $||\textbf{f}||^2=(f_1^2+f_2^2)/2$, where $f_1=x_1-\text{cos}(x_2)$ and $f_2=x_2-3\text{cos}(x_1)$. \textit{Red} point is initial choice $x^0=[-2,-2]$ and \textit{Blue} point is the solution root $(-0.684,2.32)$ with an accuracy $\varepsilon\sim10^{-11}$.} 
\end{figure} 

\section{Pseudocode}
The semi-implicit real root solver, given by Eq.~\ref{eq:6}, has been coded in MATLAB and in MAPLE. A pseudo-code is provided below. Let us first, however, discuss the basic steps as the computation proceeds. After parameter and procedure definitions and initialization with estimate $\textbf{x}^0$, there are three main steps of the iteration loop.

Firstly, for each iteration step new positions $\textbf{x}^{i+1} = \Phi(\textbf{x}^i,\textbf{A}^i)$ are computed; Eq. (5).

Secondly, an optional validity check of the iteration step is carried out if convergence is slow, i.e. if at least one member of the set $\lbrace|x_n^i-x_n^{i-1}|-|x_n^{i-1}-x_n^{i-2}|,~n=1,...,N\rbrace$ is greater than zero. Here it is controlled that convergence is quasi-monotonous and that the $\textbf{A}$ matrix is not singular or near-singular. 

The requirement $(x_n^{i}-x_n^{i+1})(x_{nk}^{i+1}-\varPhi_{nk}^{i+1})>M_c$ guarantees quasi-monotonicity. The index $k$ sweeps through $K$ equidistantly placed points in the interval $[x_n^i ,x_n^{i+1}]$. Large $K$ values result in costly function evaluations; we have found however that the value $K = 1$ is appropriate for most problems, including those of Table 1 in \citep{Scheffel:SIR}. The value $M_c = 0$ would correspond to strictly monotonous convergence, which usually is unattainable. Instead by choosing a small, negative value for $M_c$, quasi-monotonicity is allowed. 

Furthermore, if the $\textbf{A}$ matrix with components $\alpha_{mn}$ is near-singular, fatally large steps could result. We here require that all $|\alpha_{mn}|\leqslant\alpha_{max}$, where $\alpha_{max} = 2$ is a good default value. 

If either of these criteria is not satisfied for all problem dimensions, new values $R^i =(3R^i +1)/4$ are computed for at most $I_s$ subiteration steps for any relevant dimension $m$. This procedure causes the slope of the corresponding hypersurface to approach unity in the direction $m$, which supports monotonic convergence. The local gradient of the hypersurface is zero for all other dimensions. 

As a final third step, the slope parameter $R^i$ is reduced towards zero to facilitate convergence approaching second order; thus we have $R_m^{i+1}={\kappa}R_m^i$, where standard values are $R_m^{0}=0.95$ and 
$\kappa=0.5$.

The iterations proceed until the desired accuracy is achieved or until the assigned number of iterations is exceeded. It should be remarked that the use of subiterations by calling SIR-s slows down the algorithm due to additional numerical operations and should only be used when required.

A formal pseudocode for SIR is given below and in \citep{Scheffel:SIR}. In Appendices A and B, MATLAB and MAPLE codes for SIR are provided.

\section{Impact}
\noindent SIR is routinely used as a robust solver of systems of nonlinear algebraic equations in GWRM applications \citep{Scheffel:SIR}. The GWRM is a time-spectral PDE solver that has been applied to linear and nonlinear PDEs. The latter include Burger's equation, a nonlinear wave equation and chaotic equations relevant for numerical weather prediction. 

The algorithm has extended global convergence as compared to Newton's method and avoids landing on local minima, being problematic for Newton's method using linesearch. Thus SIR has a potential of being widely used in a number of computational physics areas. 

\section{Conclusion}
\noindent SIR is a recently developed solver for general systems of nonlinear algebraic equations. Global convergence is often superior to that of Newton's method and Newton's method with linesearch, as shown in an example. SIR is also simple to code; compact MATLAB and MAPLE codes are included as appendices.

\section*{References}
\bibliographystyle{elsarticle-harv} 
\bibliography{mybib}

\begin{thebibliography}{6}
\expandafter\ifx\csname natexlab\endcsname\relax\def\natexlab#1{#1}\fi
\expandafter\ifx\csname url\endcsname\relax
  \def\url#1{\texttt{#1}}\fi
\expandafter\ifx\csname urlprefix\endcsname\relax\def\urlprefix{URL }\fi

\bibitem[{Gottlieb and Orszag(1977)}]{Gottlieb:1}
Gottlieb, D., Orszag, A., 1977. Numerical Analysis of Spectral Methods: Theory
  and Applications. SIAM, Philadelphia.

\bibitem[{Harned and Kerner(1985)}]{Harned:1}
Harned, D., Kerner, W., 1985. Semi-implicit method for three-dimensional
  compressible magnetohydrodynamic simulation. Journal of Computational Physics
  60, 62--75.

\bibitem[{Harned and Schnack(1986)}]{Harned:2}
Harned, D., Schnack, D.~D., 1986. Semi-implicit method for long time scale
  magnetohydrodynamic computations in three dimensions. Journal of
  Computational Physics 65, 57--70.

\bibitem[{Scheffel(2011)}]{Scheffel:GWRM1}
Scheffel, J., 2011. Time-spectral solution of initial-value problems. In:
  Partial Differential Equations: Theory, Analysis and Applications. Nova
  Science Publishers, Inc., pp. 1--49.

\bibitem[{Scheffel and H{\aa}kansson(2009)}]{Scheffel:SIR}
Scheffel, J., H{\aa}kansson, C., 2009. {Solution of systems of nonlinear
  equations, a semi-implicit approach}. Applied Numerical Mathematics.

\bibitem[{W.~H.~Press and Flannery(2007)}]{NumericalRecipes}
W.~H.~Press, S. A.~Teukolsky, W. T.~V., Flannery, B.~P., 2007. Numerical
  Recipes; The Art of Scientific Computing 3rd Edition. Cambridge University
  Press.

\end{thebibliography}


\newgeometry{margin=4cm}
\begin{algorithm*}
{\fontsize{10pt}{9pt}\selectfont
\caption{Pseudo code: Semi-implicit root solver SIR}\label{SIRalg}
\begin{algorithmic}[H]
\Procedure{SIR}{}
\BState \textbf{input:} A vector function $\varphi:\mathbb{R}^N\rightarrow\mathbb{R}^N$ and initial estimate $\textbf{x}^0\in\mathbb{R}^N$.
\BState \textbf{output:} A vector $\textbf{x}^*$, being a root of the matrix equation $\textbf{x}=\varphi(\textbf{x})$.

\BState \textbf{parameters:} 
$N$ - number of equations to be solved, 
$tol$ -- solution accuracy,
$I_{max}$ max number of iterations, 
$sub$ - flag for allowing/omitting subiterations, 
$I_{S}$ - max number of subiterations, 
$K$ - number of monotonicity check points, 
$\alpha_{\text{c}}$ - maximum allowed magnitude of $\textbf{A}$ matrix elements,
$M_c$ - parameter for monotonicity check, 
$d{\Phi}dx_0$ - initial values of ${\partial}{\Phi}_i/{\partial}x_i$,
$R_{fac}$ - factor controlling ${\partial}{\Phi}_i/{\partial}x_i$ at each iteration. 
It is cheaper to solve $\textbf{R}=(\textbf{A}-\textbf{I})\textbf{J}+\textbf{I}$ for $\textbf{A}$ rather than computing $\textbf{J}^{-1}$ as below. 
\Statex
\State $conv=|\textbf{x}^i_n-\textbf{x}^{i-1}_n|-|\textbf{x}^{i-1}_n-\textbf{x}^{i-2}|$
\State $\textbf{R}:=d{\Phi}dx_0$, $\textbf{I}:=\text{Identity matrix}$
\State $\textbf{J}=\lbrace\partial(x_m-\varphi_m(\textbf{x}))/\partial{x_n}\rbrace$

\For {$i=1$ to $I_{max}$}
\State $\textbf{x} = \textbf{x}^0$ 
\State $\textbf{A}=\textbf{I}+(\textbf{R}-\textbf{I})\textbf{J}^{-1}$ 
\State $\textbf{x}^1=\textbf{A}(\textbf{x}-\varphi)+\varphi$

\If {$\text{sub} ~ \textbf{and} ~ \text{max}(conv)>0$} 
\For {$j=1$ to $I_{S}$}
\State $S_1=$ subiterated sequence for monotonicity test (see text)
\State $S_2=$ sequence of $|A_{nj}|$ for all dimensions n 
\If {$\text{max}(S_2) < a_c  ~ \textbf{and} ~ \text{min}(S_1) \geq M_c$} \Return false
\EndIf
\State $R_n=(3R_n+1)/4$
\State $\textbf{x} = \textbf{x}^0$ 
\State $\textbf{A}=\textbf{I}+(\textbf{R}-\textbf{I})\textbf{J}^{-1}$
\State $\textbf{x}^1=\textbf{A}(\textbf{x}-\varphi)+\varphi$
\EndFor
\EndIf
\State $\textbf{R}=R_{fac}\textbf{R}$
\State $\varepsilon=\sum^N_{n=1}|x^1_n-x^0_n|/N$
\State $\textbf{x}^0=\textbf{x}^1$
\If {$\varepsilon < \text{tol}$} \Return false
\EndIf
\EndFor
\State $\textbf{x}^*=\textbf{x}$
\EndProcedure
\end{algorithmic}}
\end{algorithm*}
\newgeometry{margin=2.5cm}

\appendix
\pagebreak
\section{Matlab code}
\label{B}
\begin{lstlisting}
function U = SIR(phi,x0,sub)

%%%%%%%%%%%%%%%%
%% Parameters %%
%%%%%%%%%%%%%%%%
if sub==0
    % Default values
    Rfac=0.5;      % Reduction of R at each iteration
    dPdx=0.95;     % Initial values of R
else
    % Values when subiterating
    Rfac=0.8;      % Reduction of R at each iteration
    dPdx=0.9999;   % Initial values of R
end

N=numel(x0);     % Number of equations
tol=1e-8;        % Solution accuracy
imax=100;         % Maximum number of iterations
Js=1000;           % Maximum number of subiterations
a_c=2;           % Critical magnitude for alpha
S1_min=-5e-2;    % Parameter for monotonicity check

%%%%%%%%%%%%%%%%%%%%
%% Initialization %%
%%%%%%%%%%%%%%%%%%%%

x0 = x0(:);             % Turn x0 into column vector
xold = zeros(size(x0)); % Store the old x0
R = dPdx*ones(N,1);     % R0

J = @(x)jacobian(phi,x,N);

%%%%%%%%%%%%%%%%%%%%
%% Iteration Loop %%
%%%%%%%%%%%%%%%%%%%%
for n = 1:imax

    % --------------------
    % I: Compute new alpha
    % --------------------

    x    = x0;            % Initial guess
    phi0 = (phi(x))';     % Evaluate initial guess
    J0   = J(x);          % Evaluate center diff if no Jacobian
    
    % Test for singularity
    % Convert to full matrix, in case J is sparse.
    condJ = rcond(full(J0));  
    if condJ <= eps
        error('Jacobian singular in SIR');
    end

    % Compute R-I
    RI0 = RI_proc(R,N);
    % Compute alpha
    alpha = alpha_proc(RI0,J0,N);
    % --------------
    % II: Compute x1
    % --------------
    x1 = Phi_proc(x0,phi0,alpha);

    %%%%%%%%%%%%%%%%%%%
    %% Subiterations %%
    %%%%%%%%%%%%%%%%%%%
    % Only subiterate if step length is increasing
    if sub && max(abs(double(x1-x0)) - abs(double(x0-xold))) > 0
        mon = x0-x1;

        % -----------------------
        % V: Check validity of x1
        % -----------------------
        for j=1:Js
            % Evaluate phi(x1)
            x = x1;            % initial guess
            phi1 = (phi(x))';  % evaluate initial guess

            S1 = mon .* (x1-Phi_proc(x1,phi1,alpha));
            S2 = max(abs(alpha),[],2);
           
            % Perform all comparisons at once, 
            % store in boolean vector ifR
            % A particular row of ifR will be true if subiteration is
            % required in that dimension.
            ifR = (S2 >= a_c | S1 < S1_min);

            % Break if no subiterations are needed
            if max(ifR) == 0
                break;
            end

            % Increase R towards I
            for i=1:N
                % Increase i:th dimension
                if ifR(i) == 1
                    R(i) = (3*R(i)+1)*0.25;
                end
            end
            % Update RI
            RI0 = RI_proc(R,N);

            % Recompute alpha
            alpha = alpha_proc(RI0,J0,N);
            x1 = Phi_proc(x0,phi0,alpha);
        end
        % -----------------------
        % Subiterations end here:
        % -----------------------
    end

    % ------------------------
    % Accuracy test and update
    % ------------------------
    eps_acc = mean(abs(x1-x0));
    xold = x0; % Backup x0 so that it can be used to compare step lengths.
    x0 = x1;
    if eps_acc < tol
        break
    end

    % ----------
    % Decrease R
    % ----------
    % This was moved to the end of the iteration step, rather than being
    % performed at the beginning of all steps but the first. This removes
    % the need to test if n>1 at each step.
    R = Rfac*R;
end
U = x0';

%%%%%%%%%%%%%%%
%% Functions %%
%%%%%%%%%%%%%%%
function alpha = alpha_proc(RI,J,N)
alpha = eye(N)+RI/J;

function Phi = Phi_proc(x,phi,alpha)
% Computes Phi (x and phi are allowed to be matrices)
Phi = alpha*(x-phi)+phi;

function RI = RI_proc(R,N)
% Computes RI
RI = zeros(N,N);
for i=1:N
    RI(i,i) = R(i)-1;
end

function J = jacobian(phi,x,N)
h=eps^(1/3);       % Finite difference length
Nf=numel(phi(x));  % Degrees of freedrom
J=zeros(N,Nf);     % Allocate J
for i=1:N
    dx=zeros(N,1); dx(i)=dx(i)+h;
    J(:,i)=(phi(x+dx)-phi(x-dx))/h/2; % Center difference
end

% If J is a cell array, we extract its component parts
if iscell(J)
    r = double(J{1});
	c = double(J{2});
	J = J{3};
	sparse_def = true; % Flag sparse version
else
	sparse_def = false; % Flag dense version
end
 
% Transform J0 into I-J0
if sparse_def
	% Sparse version
	J = sparse(r,c,J,N,N);
	J = speye(N)-J;
else
    % Dense version
	J = eye(N)-J;
end


\end{lstlisting}

\pagebreak
\section{Maple code}
\subsection{SIR}
\label{C}
\begin{lstlisting}[language=Maple]
SIR:=proc(N,phi,X,x,J,ITmax,Rfac,dPdx,sub,tol,mA)
global Phi,alpha,RI,R,E,f,nsave,U,eps:
local Js,a_c,S1_min,Phi_proc,Phi_one_proc,J_proc,i,j,k,K,m,n,alpha_proc,xold,x0,x1,mon,conv,x0_sav,xt,Phi_t,S1,S2,ifR,xv:

#-----------------#
# Root solver SIR #
#-----------------#
# This Maple code finds a solution to the system of nonlinear equations f = x - phi(x) = 0 from an initial guess solution vector X. A semi-implicit approach is used, being a generalization of Newton's method with improved global convergence characteristics. Newton's method is obtained by setting the parameter dPdx = 0. For details, see J. Scheffel and C. Hakansson, Appl. Numer. Math. 59(2009)2430.  

#------------------#
# Input parameters #
#------------------#
# N     - total number of equations
# phi   - functions in fixed point equations x = phi(x)
# X     - initial guess for
# x     - solution vector
# J     - Jacobian of f(x)=x-phi(x)
# ITmax - maximum number of iterations
# Rfac  - reduction of R at each iteration (rec = 0.5)
# dPdx  - dPhi/dx = R at first iteration (rec = 0.95)
# sub   - for subiteration set =1 (rec=0)
# tol   - solution accuracy
# mA    - number of iterations that A matrix is recomputed (Nonlinear equations: use ITmax value for default or try numbers like 5 to increase efficiency. Linear equations: since A is constant, use 0 when calling SIR more than once)

Js :=20:         # Maximum number of sub-iterations (rec = 20)   
K  :=1:          # Number of monotonicity check points (rec = 1)        
a_c:=2.0:        # Maximum allowed magnitude of alpha element (rec = 2.0)     
S1_min:=-5.0e-2: # Parameter for monotonicity check

#--Vectors and Matrices.--#
unassign('x','R'):
xv := Vector(N, symbol = x):
E  := IdentityMatrix(N,storage=sparse):
RI := DiagonalMatrix(Vector(N,symbol=R)-Vector(N,1));

#--Procedures.--#
# Semi-implicit function Phi(x;a) 
Phi_proc:=proc()            
global Phi:
Phi:=map(eval,alpha.(xv-phi)+phi):
end:

# Evaluates single row of Phi(x;a)  
Phi_one_proc:=proc(k,x)     
local m:
global Phi1:
Phi1:=add(alpha[k,m]*(x[m]-phi[m]),m=1..N)+phi[k]:
end:

# Semi-implicit parameters alpha[i,j] 
alpha_proc:=proc()          
global alpha:
alpha:=E+Transpose(LinearSolve(Transpose(J),Transpose(RI))):
end:

#--Initialization.--#
x   :=X:   # Initial guess for root x = x0                   
xold:=     Array(1..N)  :
R   :=dPdx*Array(1..N,1):
eps :=1.0: # Dummy value for accuracy test            

#--Iteration loop.--#
for n from 1 to ITmax while eps > tol do

#--First: compute new alpha matrix and iterate Phi.--#
if n<=mA then alpha_proc() fi:
Phi_proc():

#--Second: assign new x positions.--#
for i from 1 to N do
  x0[i]  := x[i]:
  x[i]   := Phi[i]:
  x1[i]  := x[i]:
  mon[i] := x0[i]-x1[i]:
od:

# Third: for sub=1; subiterate and make sure that valid step is taken.
# Valid step: 1) convergence should be monotonous  and
#             2) near-singular values of alpha should be avoided.
# Invalid step remedy; respective R[i] are adjusted towards 1 in # up to Js sub-iterations. 

if sub=1 then
if max(seq(abs(x1[i]-x0[i])-abs(x0[i]-xold[i]),i=1..N))>0 then print('inside'):
for m from 1 to Js do

  for i from 1 to N do
    x0_sav:=x0[i]:
    for k from 1 to K do xt[k]:= x0[i] + (k/K)*(x[i]-x0[i]): od:
    for k from 1 to K do     
      x0[i]:=xt[k]:
      Phi_one_proc(i,x0):
      Phi_t[k]:=Phi1: 
    od:  
    x0[i]:=x0_sav:
    S1:=seq(mon[i]*(xt[k]-Phi_t[k]),k=1..K):
    S2:=seq(abs(alpha[i,j]),j=1..N):
    if max(S2) < a_c and min(S1) >= S1_min then ifR[i]:=0 else ifR[i]:=1: fi:
  od:

  if max(seq(ifR[j],j=1..N)) = 0 then break: fi:

  for i from 1 to N do
    if ifR[i] = 1 then R[i]:=(3*R[i]+1.0)/4.0: fi:
    x[i]:=x0[i]:
  od:

  if n<=mA then alpha_proc() fi:
  Phi_proc():
  for i from 1 to N do
  x[i] :=Phi[i]:
  x1[i]:=x[i]:
  od:
od:
fi:
fi:

eps:=add(abs(x1[ie]-x0[ie]),ie=1..N)/N:
if n=ITmax and eps > tol then 
print(SIR_does_not_converge_wrt_tol_for_this_ITmax):
else
print('eps',eps,'at_iteration',n):
fi:

#--Fourth: second order convergence means letting R values tend to zero. Save also iteration data for each x component.--#
for i from 1 to N do
  R[i]   :=Rfac*R[i]:
  xold[i]:=x0[i]:
  U[i,n]:=x[i]: if n=1 then U[i,0]:=xold[i] fi:
od:

od:         # End of iteration loop

#--Fifth: save the number of iterations and the new X vector (=x). Set x free.--#
nsave:=n:
for i from 1 to N do X[i]:=x[i]: od:
unassign('x'):

end:
\end{lstlisting}
\subsection{Compute Jacobian}
\begin{lstlisting}[language=Maple]
#-----------------------------------#
#         Create Jacobian J         #
# by differentiating f(x)=x-phi(x). #
#-----------------------------------#

J_proc:=proc(Jflag,m1,m2)
global J,f:
local  i,j,k:
# The Jacobian can be obtained in three ways (provide Jflag on input):
# Jflag=1 - rather slow: based on definition and Maple differentiation: 
if Jflag=1 then
  J:=Matrix(1..N,1..N):
  for i from 1 to N do
  f[i]:=x[i]-phi[i]:
  for j from 1 to N do
  J[i,j]:=diff(f[i],x[j]):
  od: od:
fi:
# Jflag=2 - faster (recommended): uses sparse matrix algorithms when appropriate:
if Jflag=2 then
  J:=VectorCalculus[Jacobian]([seq(x[i]-phi[i],i=1..N)],[seq(x[j],j=1..N)]):
fi:
# Jflag=3 - really fast for large, sparse matrices (N>1500 or so). Algorithm below is for band matrix where parameters m1 and m2 stand for the number of subdiagonals and superdiagonals respectively: 
if Jflag=3 then
  J:=Matrix(1..N,1..N):
  for k from 0 to m1 do
  for i from m1+1-k to N do
  f[i]:=x[i]-phi[i]:
  J[i,i-m1+k]:=diff(f[i],x[i-m1+k]):
  od: od:
  for k from 0 to m2-1 do
  for j from m2+1-k to N do
  f[j-m2+k]:=x[j-m2+k]-phi[j-m2+k]:
  J[j-m2+k,j]:=diff(f[j-m2+k],x[j]):
  od: od:
fi:
end:


\end{lstlisting}




\end{document}